\documentstyle[amssymb,aps,preprint]{revtex}

\tightenlines
\begin{document}
\title{Exact-Exchange Density Functional Theory applied to a strongly inhomogeneous
electron gas}
\author{S. Rigamonti, F. A. Reboredo$^{*}$, and C. R. Proetto}
\address{Comisi\'{o}n Nacional de Energ\'{\i}a At\'{o}mica\\
Centro At\'{o}mico Bariloche and Instituto Balseiro\\
8400 Bariloche, Argentina\\
$^{*}$Lawrence Livermore National Laboratory, Livermore, California 94550,
USA}
\maketitle

\begin{abstract}
A recently developed quasi two-dimensional exact-exchange formalism within
the framework of Density Functional Theory has been applied to a strongly
inhomogeneous interacting electron gas, and the results were compared with
state-of-the-art Variational Quantum Monte Carlo (VMC) numerical simulations
for a three-dimensional electron gas under a strong external potential. The
VMC results, extremely demanding from the computational point of view, could
be considered as a benchmark for the present theory. We observe a remarkable
qualitative and quantitative agreement between both methods from the
comparison of the exchange-hole densities, exchange-energy densities, and
total exchange-energies per particle. This agreement is increasingly
improved with the strength of the external potential when the electron gas
becomes quasi-two-dimensional.
\end{abstract}

\newpage

\narrowtext 

The search for accurate calculation schemes within the framework of Density
Functional Theory (DFT) is of paramount importance in atomic, molecular, and
solid state physics.\cite{parr} While DFT maps the problem of interacting
inhomogeneous electron systems to a problem of effectively non-interacting
inhomogeneous electrons, it provides no help in the determination of the
exchange-correlation component of the total energy functional.\cite
{hohen,kohn} Largely, this obstacle has been overcome by using the so-called
Local Density Approximation (LDA), which is based in the approximate
equivalence in the long-wavelength limit between the inhomogeneous and
homogeneous interacting electron gases, with the latter taken as a reference
system.\cite{lda} The LDA (and generalized gradient corrections to the LDA,
as the GGA) has been widely employed, however, even for strongly
inhomogeneous systems, such as atoms and molecules, with surprisingly good
results.\cite{williams} The application to strongly inhomogeneous solid
state systems has been equally successful,\cite{ando} but some warnings
raised quite recently when applied to two-dimensional electron gases (2DEG),%
\cite{kim} of the type formed at the interface between two semiconductors
such as {\rm GaAs/Al}$_{x}${\rm Ga}$_{1-x}${\rm As.}\cite{bastard}{\rm \ }%
Recently a study by Nekovee, Foulkes, and Needs (NFN) using Variational
Quantum Monte Carlo (VMC) numerical simulations showed significant failures
of the LDA and GGA in the description of the exchange hole of a 3D electron
gas under the effect of a strong periodic potential.\cite{nfn} Since plane
wave expansions are a routine approach in DFT {\it ab-initio} codes NFN
results call for methods that could describe accurately the electron gas in
regions where the chemical bond is formed.

In this paper we present the first application of a recently developed quasi
two-dimensional exact-exchange DFT (XX-DFT) method to a strongly
inhomogeneous ``metallic'' system; we have found that this local exact
exchange theory overcomes the problems presented by LDA and GGA. Previous
application of the XX-DFT formalism includes atomic and molecular systems,%
\cite{xxatoms} semiconductors (insulators),\cite{xxsemiconductors} and quasi
2DEG's.\cite{xx2deg} The method is distinguished from LDA, by avoiding the
approximation of taking the exchange energy of the inhomogeneous electron
gas as given by the exchange energy of the homogeneous electron gas,
evaluated at the same (local) value of the density of the inhomogeneous
system. Instead, the XX-DFT formalism uses the fact that the exchange energy
is an explicit function of the {\it occupied} orbitals,\cite{sahni} and
exploit this to derive a local potential that minimizes a Kohn-Sham energy
which includes the exchange energy exactly.\cite{xxtheory} This elaborate
treatment of the exchange potential, beyond the standard LDA-DFT scheme, was
shown to be crucial in providing a theoretical explanation of striking
experimental results in asymmetric semiconductor quantum wells.\cite{xx2deg}
These experiments and theoretical analysis suggest that the occupancy of a
quantum well subbands proceed in an abrupt way, mediated by inter-subband
exchange, the explanation being quite natural in the XX-DFT scheme.

The system under study is shown schematically in Fig. 1, and has been
motivated by the recent analysis of a similar system by NFN using VMC
numerical simulations;\cite{nfn} the same system has been also studied by
Rushton, Tozer, and Clark,\cite{rushton} using the so-called Weighted
Density Approximation (WDA), that treats both exchange and correlation in an
approximate, but non-local, way.\cite{wda} Our model consists of a
cosenoidal double well external potential $V_{ext}(z)=V_{0}\cos (qz),$ which
strongly modulates the electronic density along the $z$ direction, but keeps
the translational symmetry in the $x-y$ plane; the system is ultimately
confined by two infinite barriers located at $z=\pm $ $(d+\lambda ).$ The
charge neutrality of the system was fulfilled by means of a positive jellium
slab of uniform density $\overline{n}=3/4\pi r_{s}^{3},$ where $%
r_{s}=r_{0}/a_{0},$ with $4\pi r_{0}^{3}/3$ being the volume per electron,
and $a_{0}=0.529$ $\AA$ the Bohr radius. Following Ref. (9), $r_{s}$ was
fixed at $r_{s}=2$, while the amplitude $V_{0}$ was chosen to be $%
V_{0}=2.08\varepsilon _{F}^{0},$ where $\varepsilon _{F}^{0}$ is the Fermi
energy corresponding to $\overline{n};$ these values corresponds
approximately to those of {\rm Al}$.$ We studied systems with three
different modulations $q$ of the external potential, corresponding to $%
q_{1}=1.11k_{F}^{0},$ $q_{2}=1.55k_{F}^{0},$ $q_{3}=2.17k_{F}^{0},$ with $%
a_{0}k_{F}^{0}=(3\pi ^{2}\overline{n})^{1/3}.$ Note that in our model,
decreasing $q$ leads to an increasing $\lambda $, and consequently to
increasingly smooth potentials (as the potential barrier height $V_{0}$ is
fixed), approaching the three-dimensional uniform situation.

Before starting with calculations, it is important to realize that $\lambda
=2\pi /q\simeq 1$ $/$ $k_{F}^{0};$ this leads to a strong quantization of
the electronic degrees of freedom along $z.$ Following the terminology of
the quasi-2DEG's, we will denote as ``subbands'' each one of these quantized
discrete levels. As electrons are free to move in the $x-y$ plane, an
in-plane parabolic dispersion relation is associated to each one of these
subbands. To proceed, and exploiting the translational symmetry along the $%
x-y$ plane (area $A$), we propose as solutions of the three-dimensional
Kohn-Sham (KS) equations $\phi _{\nu {\bf k}\sigma }\left( {\bf r}\right)
=\exp (i$ ${\bf k}\cdot {\bf \rho })\xi _{\nu }^{\sigma }\left( z\right) /%
\sqrt{A},$ with ${\bf k}$ the in-plane wave vector, ${\bf \rho }$ the
in-plane coordinate, $\xi _{\nu }^{\sigma }\left( z\right) $ the subband
wave function corresponding to an electron with a spin projection $\sigma $ (%
$\uparrow $ or $\downarrow $) and subband index $\nu $ $(\nu =0,1,2,...).$
Proceeding with these solutions, the KS equations take the form of effective
one-dimensional Schr\"{o}dinger-like equations\cite{units} 
\begin{equation}
\left[ -\frac{1}{2}\frac{\partial ^{2}}{\partial z^{2}}+V_{KS}\left(
z,\sigma \right) \right] \xi _{\nu }^{\sigma }\left( z\right) =\varepsilon
_{\nu }^{\sigma }\xi _{\nu }^{\sigma }\left( z\right) ,
\end{equation}
with $V_{KS}(z,\sigma )$ being the local KS spin-dependent potential, and $%
\varepsilon _{\nu }^{\sigma }$ the eigenvalues$.$ $V_{KS}(z,\sigma )$ is the
sum of several contributions 
\begin{equation}
V_{KS}(z,\sigma )=V_{ext}(z)+V_{H}(z)+V_{x}^{i}(z,\sigma )+V_{c}(z,\sigma ),
\end{equation}
with $V_{H},$ $V_{x}^{i},$ and $V_{c}$ being the Hartree, exchange, and
correlation contributions, respectively. The index $i$ stands for the
exact-exchange $(i=XX)$ or Local Density Approximation $(i=LDA).$ The
Hartree potential is given by its classical expression, 
\begin{equation}
V_{H}(z)=-2\pi \int\limits_{-(\lambda +d)}^{\lambda +d}dz^{\prime }\left|
z-z^{\prime }\right| \left[ n(z^{\prime })-n_{jell}(z^{\prime })\right] ,
\end{equation}
with $n_{jell}(z)=\overline{n}$ $\theta (\lambda +z)\theta (\lambda -z),$
and $\theta (x)$ being the step function $(\theta (x)=1$ if $x>0,$ $\theta
(x)=0$ if $x<0$ $)$. For $V_{c}(z,\sigma )$ we follow the DFT prescription
and define it through the functional derivative with respect to the density
of the correlation contribution $E_{c}$ to the total energy, 
\begin{equation}
V_{c}(z,\sigma )=\frac{\delta E_{c}}{\delta n(z,\sigma )},
\end{equation}
and use LDA for its practical evaluation, using a recent parametrization of
the correlation energy per particle for the interacting homogeneous electron
gas by Gori-Giorgi, Sacchetti, and Bachelet.\cite{gsb} In equations above, $%
n(z)$ is the zero-temperature {\it three-dimensional} electron density 
\begin{equation}
n(z){}\!=\!{}\sum_{\sigma }n(z,\sigma )=\!\frac{1}{4\pi }\!\!\!\!\text{ }%
\sum_{\nu \sigma }\!\!\!\!\text{ }(k_{F}^{\nu \sigma })^{2}\left| \xi _{\nu
}^{\sigma }\left( z\right) \right| ^{2}\!,
\end{equation}
$n(z,\sigma )$ being the fraction of $\sigma $ polarized electrons. $\mu $
is the chemical potential (or Fermi level) of the system, which is
determined by the electrostatic and thermodynamic equilibrium with a
reservoir, and $k_{F}^{\nu \sigma }=\sqrt{2(\mu -\varepsilon _{v}^{\sigma })}%
\theta (\mu -\varepsilon _{v}^{\sigma }).$ It remains to define $%
V_{x}^{XX}(z).$ As explained above, we avoid its usual local (LDA), or
semilocal (GGA) expressions, using instead the recently developed XX-DFT
formalism.\cite{xxatoms,xxsemiconductors} As shown elsewhere,\cite{xx2deg}
the exact-exchange potential for our quasi two-dimensional geometry is
defined by the equation 
\begin{eqnarray}
V_{x}^{XX}(z,\sigma ) &=&\frac{\delta E_{x}}{\delta n\left( z,\sigma \right) 
}=A\sum_{\nu }\int dz^{\prime }\left\{ \int dz^{\prime \prime }\left[ \frac{%
\delta E_{x}}{\delta \xi _{\nu }^{\sigma }\left( z^{\prime \prime }\right) }%
\frac{\delta \xi _{\nu }^{\sigma }\left( z^{\prime \prime }\right) }{\delta
V_{KS}\left( z^{\prime },\sigma \right) }+c.c.\right] \right. + \\
&&+\left. \left[ \frac{\delta E_{x}}{\delta k_{F}^{\nu \sigma }}\frac{\delta
k_{F}^{\nu \sigma }}{\delta V_{KS}\left( z^{\prime },\sigma \right) }\right]
\right\} \frac{\delta V_{KS}\left( z^{\prime },\sigma \right) }{\delta
n\left( z,\sigma \right) }.  \nonumber
\end{eqnarray}
In equation above, the functional derivatives $\delta E_{x}/\delta \xi _{\nu
}^{\sigma }\left( z\right) $ and $\delta E_{x}/\delta k_{F}^{\nu \sigma }$
can be evaluated directly from the explicit expression for the exchange
energy $E_{x}$ in terms of $\xi _{\nu }^{\sigma }\left( z\right) $ and $%
k_{F}^{\nu \sigma }.$ $\delta \xi _{\nu }^{\sigma }\left( z\right) /\delta
V_{KS}\left( z^{\prime },\sigma \right) $ and $\delta k_{F}^{\nu \sigma
}/\delta V_{KS}\left( z,\sigma \right) $ are evaluated using first-order
perturbation theory from Eq. (1). Finally, $\chi _{\sigma }^{-1}(z,z^{\prime
})\equiv \delta V_{KS}\left( z^{\prime },\sigma \right) /\delta n\left( z%
{\bf ,}\sigma \right) $ is the inverse of the operator $\chi _{\sigma
}(z,z^{\prime })\equiv \delta n\left( z{\bf ,}\sigma \right) /\delta
V_{KS}\left( z^{\prime },\sigma \right) $ given by 
\begin{equation}
\chi _{\sigma }(z,z^{\prime })=\frac{1}{2\pi A}\sum_{\nu }^{occ}\left[
\sum_{\nu ^{\prime }(\neq \nu )}\left( k_{F}^{\nu \sigma }\right) ^{2}\frac{%
\xi _{\nu }^{\sigma }\left( z\right) \xi _{\nu ^{\prime }}^{\sigma }\left(
z\right) \xi _{\nu ^{\prime }}^{\sigma }\left( z^{\prime }\right) \xi _{\nu
}^{\sigma }\left( z^{\prime }\right) }{\varepsilon _{\nu }^{\sigma
}-\varepsilon _{\nu ^{\prime }}^{\sigma }}-\left[ \xi _{\nu }^{\sigma
}\left( z\right) \xi _{\nu }^{\sigma }\left( z^{\prime }\right) \right]
^{2}\right] .  \label{xhi}
\end{equation}
The first term in Eq. (\ref{xhi}) comes from first-order perturbation theory%
\cite{xxsemiconductors}, whereas the second term results from first-order
perturbation theory and the thermodynamic equilibrium between the electronic
system and the reservoir that fixes a common chemical potential $\mu $
allowing (in principle) the change of the number of particles.\cite{chemical}
Indeed, without the reservoir, the operator $\chi _{\sigma }(z,z^{\prime })$
in general cannot be inverted because it is singular\cite{xxsemiconductors}.
The alternative to $V_{x}^{XX}(z)$ is the widely employed $V_{x}^{LDA}(z),$
defined as 
\begin{equation}
V_{x}^{LDA}(z,\sigma )=-\left[ \frac{6n(z,\sigma )}{\pi }\right] ^{1/3}.
\end{equation}
As it is well known, Eq.(1), together with Eqs.(3), (4), and (6) for XX-DFT,
or (7) for LDA-DFT should be solved iteratively in a self-consistent way.
All results denoted as XX (LDA) in what follows have been obtained by
performing this numerical self-consistent procedure with $V_{x}^{XX}(z)$ $%
\left( V_{x}^{LDA}(z)\right) $inserted in Eq.(1), and by assuming a
paramagnetic situation for the present high-density metallic system. Using
as input the set of parameters defined above, the output of this
self-consistent calculation correspond to a situation with two (four,
counting spin) occupied subbands $(\nu =0,1)$ for the three different
modulations $q/k_{F}^{0}$.

One way to characterize the accuracy of a many-body approach, is to analyze
the exchange-correlation hole density, in terms of which the
exchange-correlation energy could be defined. As the emphasis of the present
work is on exchange effects, we will restrict our analysis mainly to the
exchange-hole density. We start from the exact expression for the
exchange-hole density,\cite{gunnarson} 
\begin{equation}
h_{x}({\bf r},{\bf r}+{\bf R})\equiv -\frac{1}{2}\frac{\left| \rho ({\bf r},%
{\bf r}+{\bf R})\right| ^{2}}{n({\bf r})},  \label{xhole}
\end{equation}
where 
\begin{equation}
\rho ({\bf r},{\bf r}+{\bf R})\equiv \sum_{\nu {\bf k}\sigma }\theta \left[
\mu -\varepsilon _{\nu }^{\sigma }({\bf k})\right] \phi _{\nu {\bf k}\sigma
}^{*}\left( {\bf r+R}\right) \phi _{\nu {\bf k}\sigma }\left( {\bf r}\right) 
\end{equation}
is the density matrix associated to our problem. $\varepsilon _{\nu
}^{\sigma }({\bf k})=\varepsilon _{\nu }^{\sigma }+(\hbar {\bf k)}^{2}/2m$
and note that $\rho ({\bf r},{\bf r})=n({\bf r}).$ The physical meaning of $%
h_{x}({\bf r},{\bf r}+{\bf R})$ is that it represents the density of the
exchange hole at point ${\bf r}+{\bf R}$ (observational point) due to the
presence of an electron (test particle) located at ${\bf r}$. Evaluating Eq.(%
\ref{xhole}) in our quasi two-dimensional geometry of Fig. 1, we obtain 
\begin{equation}
h_{x}({\bf r},{\bf r}+{\bf R})=-\frac{1}{2(\pi R_{\shortparallel })^{2}n(z)}%
\sum_{\nu ,\nu ^{\prime }}k_{F}^{\nu }k_{F}^{\nu ^{\prime }}J_{1}(k_{F}^{\nu
}R_{\shortparallel })J_{1}(k_{F}^{\nu ^{\prime }}R_{\shortparallel })\xi
_{\nu }^{*}(z+Z)\xi _{\nu }(z)\xi _{\nu ^{\prime }}(z+Z)\xi _{\nu ^{\prime
}}^{*}(z),  \label{xhole2D}
\end{equation}
where ${\bf R}=(R_{\shortparallel },Z),$ and $J_{1}(x)$ is the first-order
cylindrical Bessel function.\cite{mattsson} Taking the limit $%
R_{\shortparallel },Z\rightarrow 0,$ it is easy to check that $h_{x}({\bf r},%
{\bf r})=-n({\bf r})/2,$ and by explicit integration the fulfillment of the
exact constraint\cite{gunnarson} 
\begin{equation}
\int d{\bf R}\text{ }h_{x}({\bf r},{\bf r}+{\bf R})=-1,  \label{constraint}
\end{equation}
is immediate. In Eq.(\ref{xhole2D}), it is interesting to note, first, the
absence of the in-plane coordinate ${\bf \rho }$ corresponding to the test
particle , which result from the translational symmetry in the $x-y$ plane;
choosing ${\bf \rho }=0$ without any loss in generality, $h_{x}({\bf r},{\bf %
r}+{\bf R})$ could be equally written as $h_{x}(z;z+Z,R_{\shortparallel }).$
Second, the dependence of $h_{x}$ only on the magnitude of $%
R_{\shortparallel }$ means that the exchange hole has cylindrical symmetry
along the $z$ axis. Third, integration of $n(z)\times
h_{x}(z;z+Z,R_{\shortparallel })/R$ $\left( R=\left| {\bf R}\right| \right) $
in $z,Z,$ and $R_{\shortparallel }$ leads exactly to the definition of the
exchange energy used to derive $V_{x}^{XX}(z)$ in Eq.(6). In the limit $%
k_{F}^{\nu }R_{\shortparallel }\gg 1$ (assuming a finite occupancy of all
occupied subbands), and using the asymptotic expansion of the Bessel
function, we obtain 
\begin{eqnarray}
h_{x}({\bf r},{\bf r}+{\bf R}) &\simeq &-\frac{1}{(\pi R_{\shortparallel
})^{3}n({\bf r})}\times   \nonumber \\
&&\sum_{\nu ,\nu ^{\prime }}\sqrt{k_{F}^{\nu }k_{F}^{\nu ^{\prime }}}\cos
(k_{F}^{\nu }R_{\shortparallel }-\frac{3\pi }{4})\cos (k_{F}^{\nu ^{\prime
}}R_{\shortparallel }-\frac{3\pi }{4})\xi _{\nu }^{*}(z+Z)\xi _{\nu }(z)\xi
_{\nu ^{\prime }}(z+Z)\xi _{\nu ^{\prime }}^{*}(z).  \label{asymptotic}
\end{eqnarray}
From Eq. (\ref{asymptotic}), is evident that the asymptotic behavior of $%
h_{x}({\bf r},{\bf r}+{\bf R})$ is given by $R_{\shortparallel }^{-3}$ along
the $x-y$ plane; on the other side, the decay along $z$ is controlled by the
extent of the subband wave-functions $\xi _{\nu }(z).$ In what follows, we
will denote by $h_{x}^{XX}({\bf r},{\bf r}+{\bf R})$ the exchange-hole
density as given by Eq.(\ref{xhole2D}), evaluated with the self-consistent
subband wavefunctions $\xi _{\nu }(z)$ that are the solutions of Eq.(1) with 
$V_{x}^{XX}(z).$ On the other side, evaluation of Eq.(\ref{xhole}) in the
LDA-DFT scheme using 3D plane waves yields 
\begin{equation}
h_{x}^{LDA}({\bf r},{\bf r}+{\bf R})=-\text{ }\frac{n({\bf r})}{2}\text{ }%
F\left[ k_{F}({\bf r})R\right] ,  \label{xholeLDA}
\end{equation}
with $F(x)=9\left[ (\sin x-x\cos x)/x^{3}\right] ^{2}.$ In Eq.(\ref{xholeLDA}%
), and according to the LDA prescription $k_{F}({\bf r})=\left[ 3\pi ^{2}n(%
{\bf r})\right] ^{1/3}.$ We will denote by $h_{x}^{LDA}({\bf r},{\bf r}+{\bf %
R})$ the exchange-hole density as given by Eq.(\ref{xholeLDA}), evaluated
with the self-consistent subband wavefunctions $\xi _{\nu }(z)$ that are the
solutions of Eq.(1) with $V_{x}^{LDA}(z).$ It is easy to see that $%
F(x)\rightarrow 1$ for $x\rightarrow 0,$ while it decays as $x^{-4}$ for $%
x=k_{F}({\bf r})R\gg 1.$ This simple dependence of $F(x)$ on $x$ implies
that $h_{x}^{LDA}$ is {\it always} centered on the electron coordinate ${\bf %
r}$. Besides, it is clear that $h_{x}^{LDA}({\bf r},{\bf r}+{\bf R})$ is a
spherically symmetric function of ${\bf R},$ and centered at ${\bf r}$.

Fig. 2 allows a qualitative comparison of both holes at a density maximum,
corresponding to the electron coordinate ${\bf \rho }=0,$ $z=-\lambda /2$
being located at the left potential well of Fig. 1. For this case, both
holes are centered at the electron coordinate $.$ Note that for both upper
and lower panels, only half of the exchange hole has been represented; the
full hole could be obtained by reflection with respect to the $z$ axis. The
exact-exchange hole is contracted along the confinement direction, due to
the fact allowed above that the decay along the $x-y$ plane and along the $z$
direction are different, as seen explicitly from Eq. (\ref{asymptotic}). The
fact that $h_{x}^{XX}({\bf r},{\bf r}+{\bf R})$ has essentially zero
strength at the right well minima, simply means that the system is in the
``atomic'' or isolated well limit, with a very small amount of tunneling or
subband wave-function overlap between the two wells. It is also interesting
to note the Friedel-like oscillations of the exact-exchange and LDA holes,
each of them reflecting the different types of symmetry of the corresponding
holes.

The situation displayed in Fig. 3, corresponding to the electron coordinate $%
{\bf \rho }=0,$ $z=0$ at the density minimum, is strikingly different,
compared with Fig. 2. Here, $h_{x}^{XX}({\bf r},{\bf r}+{\bf R})$ attains
its maximum strength when the observational coordinate $Z$ coincides with
the density maximum $(z=\pm \lambda /2).$ This is the physically correct
behavior that one expects for the exchange-hole density, and that can easily
be obtained from Eq.(\ref{xhole2D}): Restricting for simplicity the analysis
to the case where only one subband $(\nu =0)$ is occupied, this equation
simplifies to 
\begin{equation}
h_{x}^{XX}({\bf r},{\bf r}+{\bf R})=-\frac{\left[
J_{1}(k_{F}^{0}R_{\shortparallel })\right] ^{2}}{\pi R_{\shortparallel }^{2}}%
\left[ \xi _{0}(z+Z)\right] ^{2},
\end{equation}
which clearly attains its maximum strength, as a function of $Z,$ when the
argument of the subband wave function $(z+Z)$ coincides with a density
maximum. This non-local behavior of the exact-exchange hole is clearly
preserved when more than one subband is occupied (two subbands are occupied
for the situation displayed in Figs. 2, 3, and 4). For $z=\pm \lambda /2$
(Fig. 2) this happens for $Z=0,$ while for $z=0,$ this happens for $Z=\pm
\lambda /2.$ The LDA exchange hole, on the other side, is centered at the
electron coordinate $z.$ Attention should be paid also to the quantitative
values of $h_{x}^{XX}({\bf r},{\bf r}+{\bf R})$ and $h_{x}^{LDA}({\bf r},%
{\bf r}+{\bf R})$ in Fig. 3. In the exact-exchange case, $h_{x}^{XX}$ has a
maximum strength value of about half of its value in Fig. 2, which is
reasonable keeping in mind the constraint given by Eq.(\ref{constraint}). On
the other side, $h_{x}^{LDA}$ has a maximum strength of about two orders of
magnitude smaller than $h_{x}^{XX},$ and accordingly it has an ``anomalous''
large extent in real space in order of satisfy Eq. (\ref{constraint}). Note
also that the extent in real space for $h_{x}^{XX}({\bf r},{\bf r}+{\bf R})$
is similar in Figs. 2 and 3, as it should be, as in this case and for both
situations the size of the hole is dictated by the accumulation points of
electronic density.

We show in Fig. 4 the exchange-energy densities differences, defined as
follows: $\Delta e_{x}^{VMC}(z)=e_{x}^{XX}(z)-e_{x}^{VMC}(z)$ (full thick
line), and $\Delta e_{x}^{LDA}(z)=e_{x}^{XX}(z)-e_{x}^{LDA}(z)$ (dashed
thick line). Exchange-energy densities are obtained from 
\begin{equation}
e_{x}^{i}({\bf r})=\frac{1}{2}\int \frac{d{\bf R}}{R}\text{ }n({\bf r})\text{
}h_{x}^{i}({\bf r,r+R})
\end{equation}
with $i=XX,LDA.$ The full thin line corresponds to the XX density $n^{XX}(z)$%
, the dotted thin line to the VMC density $n^{VMC}(z)$. We will discuss
first the comparison between XX and VMC calculations. While in principle
both methods give an exact treatment to the exchange-energy, and accordingly
they should coincide, in practice there are three possible sources for the
small differences observed in Fig. 4: $i)$ while our double quantum well
system is a very good approximation to the periodic system studied by NFN,
the equivalence is not perfect from the ``model'' point of view,\cite{model} 
$ii)$ even if both models were identical, for the practical implementation
of the VMC calculation, the external potential $V_{ext}(z)$ used by NFN is
not identical of ours (although it should be close to it), and $iii)$ even
if both the model {\it and} the external potential were identical, still we
are treating correlation at the LDA level, while NFN treats correlation with
VMC accuracy, presumably a much better approximation than LDA, and
definitively quite different. As a consequence of these three factors, small
differences between the two different densities profiles result, which in
turn translate to the small differences for the exchange-energy densities
observed in Fig. 4. It is important to realize, however, that points $i$ and 
$ii$ above are just consequence of the different scheme of calculations,
while point $iii$ is the important and intrinsic difference between XX-DFT
and VMC. It is also interesting to note that the differences between XX and
VMC are mainly concentrated in regions of high density, as expected;
besides, differences increase from top (small $q$) to bottom (large $q$), as
both models increasingly disagree in the large $q$ limit. While from the
results presented in Table I for the corresponding integrated magnitudes the
difference between XX and VMC is smallest for the intermediate case $%
q/k_{F}^{0}=1.55,$ from Fig. 4 it is clear that this is due to an
(accidental) error cancellation. Based in the almost perfect agreement
between $n^{XX}(z)$ and $n^{VMC}(z)$ displayed in Fig. 4 for $%
q/k_{F}^{0}=1.11,$ it is clear that our modelization of the periodic system
studied by NFN is at best in this quasi two-dimensional limit (see below).
It is then plausible to assume that the remaining discrepancy between XX and
VMC is mainly due to the different treatment of correlation in both cases.
In this context, it is interesting to point out that in our case, the
correlation energy per particle is typically 15 \% of the
exchange-correlation energy per particle. The comparison between XX and LDA
exchange-energy densities shows large local deviations between the two
methods, the amplitude of the differences being about a factor of two
greater than between XX and VMC.

The situation is similar for the differences in the integrated energies
presented in Table I, the differences between XX and VMC being much smaller
than between XX and LDA for $q/k_{F}^{0}=1.11$ and $1.57.$ The difference
between XX and LDA exchange energies (per particle) is smaller than between
XX and VMC for $q/k_{F}^{0}=2.17.$ As can be seen from Fig. 4, this is in
part due to a large error cancellation of the LDA results, which, being
non-systematic is difficult to control, and also due to the fact that as $q$
increases, our model and the NFN model move towards different limits (see
below).

We show in Fig. 5 the exchange energy per particle $\varepsilon _{x}$,
calculated in different approximations, as a function of the potential
modulation $q/k_{F}^{0}.$ In order to place these results in perspective, it
is important to note that the exchange energy per particle for the
homogeneous 3D electron gas is 
\begin{equation}
\varepsilon _{x}^{3D}(r_{s})=-\frac{3}{4\pi }\left( \frac{9\pi }{4}\right)
^{1/3}\frac{1}{r_{s}},
\end{equation}
while for a 2D homogeneous electron gas it is 
\begin{equation}
\varepsilon _{x}^{2D}(r_{s})=-\frac{4\sqrt{2}}{3\pi }\frac{1}{r_{s}}.
\end{equation}
For $r_{s}=2,$ we obtain $\varepsilon _{x}^{3D}(r_{s}=2)\simeq -0.23,$ $%
\varepsilon _{x}^{2D}(r_{s}=2)\simeq -0.30.$ For $q/k_{F}^{0}\rightarrow 0,$
the system resembles a three-dimensional uniform electron gas, which
explains the decreasing (in absolute value) of the exchange energy for $%
q/k_{F}^{0}\lesssim 1.$ While our XX-DFT formalism is ideally suited for the
study of strongly inhomogeneous systems, it is numerically ineficcient for
reaching the 3D limit, when a large number of subbands become occupied and
the (numerical) calculation becomes increasingly cumbersome.\cite{3d} The
tendency of the exchange energy towards the homogeneous 3D limit $%
\varepsilon _{x}^{3D}(2)\simeq -0.23$ is however clearly displayed in Fig.
5. It is also interesting to note how LDA becomes an increasingly better
approximation to XX for $q/k_{F}^{0}\rightarrow 0,$ as one expects. The
opposite limit $q/k_{F}^{0}\gtrsim 1$ is also worth analyzing. In this case,
the periodic model studied by NFN becomes increasingly three-dimensional, as
the fixed-height barriers becomes thinner and electrons are increasingly
able of tunnel through them. Our model, on the other side, in the limit $%
q/k_{F}^{0}\gtrsim 1$, and as a consequence of the constraint $\lambda =2\pi
/q$, departs from the NFN model and becomes more similar to an infinite
barrier quantum well of width $2d$ with a positively charged layer at its
center. The different limits reached by the model of NFN and ours explains
the progressive disagreement between XX and VMC which is seen for $%
q/k_{F}^{0}\gtrsim 2.$ Besides these two limiting features, clearly the more
interesting region is the intermediate regime about $q/k_{F}^{0}\simeq 1,$
where the inhomogeneity reaches its maximum strength. The proximity of the $%
\varepsilon _{x}^{XX}$ and $\varepsilon _{x}^{VMC}$ to the $2D$ value $%
\varepsilon _{x}^{2D}(2)\simeq -0.30$ tells us that for $q/k_{F}^{0}\simeq 1$
the system is quite close to the strict two-dimensional limit. Quite good
agreement is observed between XX and VMC in this strongly inhomogeneous
regime, as according to Table I, the percentage error between both
calculations is about 0.65 \%. This should be compared with the
corresponding error between XX and (our) LDA, that reaches a value of about
2.4 \% in the same quasi two-dimensional regime, almost four times higher.

In summary, we have checked the accuracy of the exact-exchange density
functional method as applied to a strongly inhomogeneous interacting
electron gas, and compared with the best available results. We have analyzed
the exact-exchange hole density, and found that it is strongly non-local
when evaluated in regions of low-density. From the comparison with the
results of Variational Quantum Monte Carlo, we found a remaining discrepancy
of about 0.65 \% between the exact-exchange and VMC exchange contributions,
that we attribute to the different treatment of correlation in both
approaches. Exact-Exchange Density Functional Theory works at best in the
worse case for LDA, that is, when the strongly inhomogeneous electron gas is
closer to the quasi two-dimensional limit. Our results motivate the urge for
better and more accurate correlation functionals, and work in this direction
is in progress.

The authors are very grateful to the authors of Ref.(9), who provides the
data on their VMC calculations displayed in Fig. 4 prior to publication and
for useful comments, and to K. Hallberg for a critical reading of the
manuscript. We also benefited from the code provided by the authors of
Ref.(18) for the calculation of the correlation energy per particle. S.R.
acknowledges support from the Instituto Balseiro of S. C. de Bariloche, and
F.A.R. and C.R.P. are indebted to CONICET of Argentina for financial support.

\begin{figure}[tbp]
\caption{ Schematic representation of the external potential $%
V_{ext}(z)=V_{0}\cos (qz).$ The wave-length of the modulation is $\lambda
=2\pi /q,$ and the amplitude $V_{0}=2.08$ $\varepsilon _{F}^{0}.$ $d=2$ $%
a_{0}\simeq 1$ A is the distance between the end of the jellium slab (shaded
region) and the infinite potential barriers located at $z=\pm (d+\lambda ).$}
\end{figure}

\begin{figure}[tbp]
\caption{ Exchange-hole density at a density maximum ( $z=-\lambda /2$ in
Fig. 1), for $q/k_{F}^{0}=1.11$. Upper panel: $h_{x}^{XX}({\bf r},{\bf r}+%
{\bf R}).$ Lower panel: $h_{x}^{LDA}({\bf r},{\bf r}+{\bf R}).$}
\end{figure}
\begin{figure}[tbp]
\caption{ Exchange-hole density at a density minimum ( $z=0$ in Fig. 1), for 
$q/k_{F}^{0}=1.11$. Upper panel: $h_{x}^{XX}({\bf r},{\bf r}+{\bf R}).$
Lower panel: $h_{x}^{LDA}({\bf r},{\bf r}+{\bf R}).$ It is important to
realize that the numerical value of $h_{x}^{LDA}$ at its point of {\it %
maximum} strength coincides with the value of $h_{x}^{XX}$ at its point of 
{\it minimum} strenght, marked with a square full dot. }
\end{figure}

\begin{figure}[tbp]
\caption{ Left-scale, full (dashed) thick line: Difference between XX and
VMC (LDA) exchange-energy densities $\Delta
e_{x}^{VMC}(z)=e_{x}^{XX}(z)-e_{x}^{VMC}(z)$ $\left(
e_{x}^{XX}(z)-e_{x}^{LDA}(z)\right) ,$ versus position in the system, and
for the three modulations. Right-scale: exact-exchange (full thin line) and
VMC (dotted thin line) electronic densities $n^{XX}(z)$ and $n^{VMC}(z)$,
respectively; note that both sets of densities are almost indistinguishable
except for $q/k_{F}^{0}=2.17$and $\left| z\right| \simeq \lambda $. Data
corresponding to $e_{x}^{VMC}(z)$ and $n^{VMC}(z)$ provided by the authors
of Ref.(23). }
\end{figure}
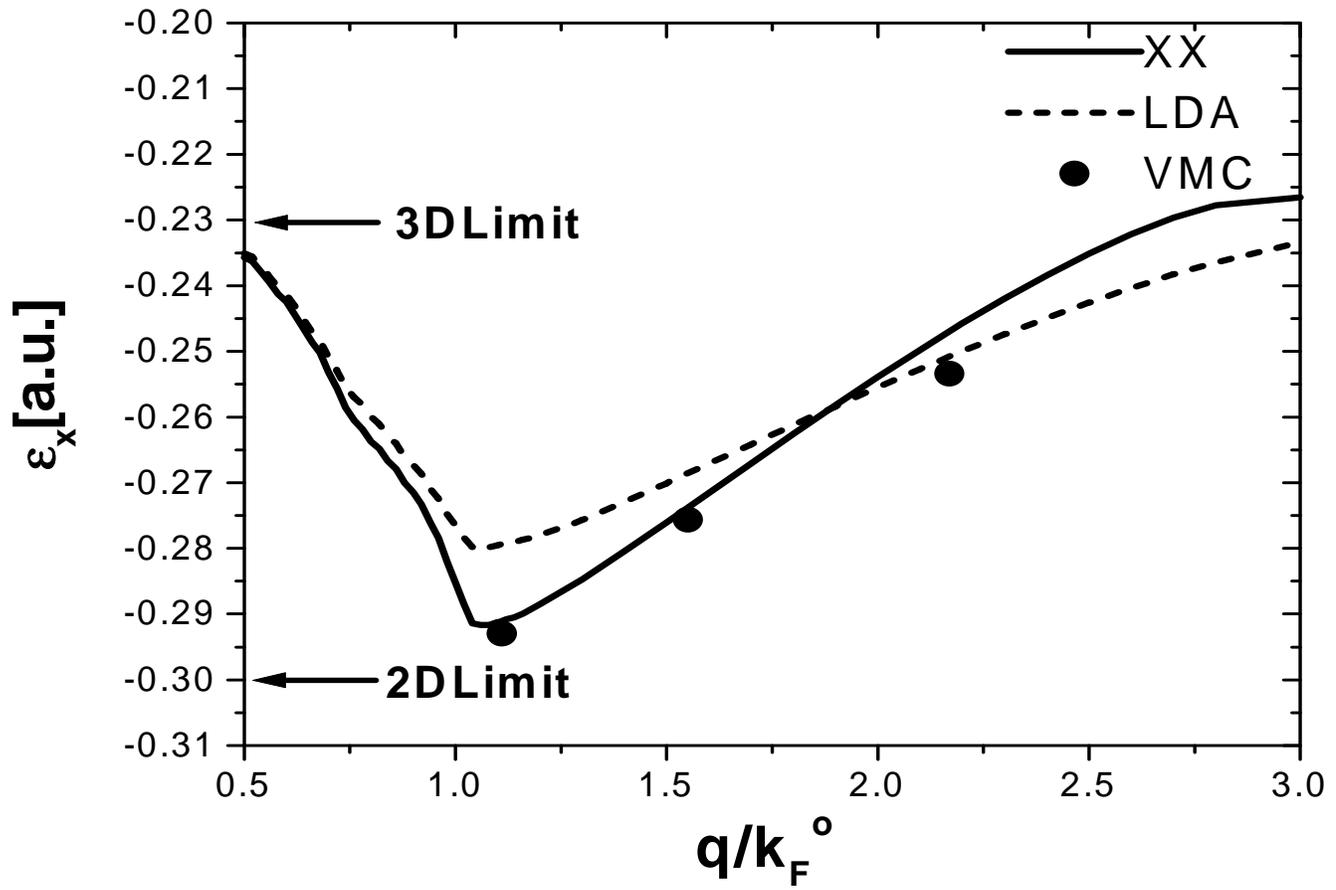
\begin{figure}[tbp]
\caption{ Exchange-energy per particle as a function of the potential
modulation. Full and dashed lines are our XX $(\varepsilon _{x}^{XX})$ and
LDA $(\varepsilon _{x}^{LDA})$ results, respectively. Discrete points are
VMC data from Ref. 9.}
\end{figure}

\begin{table}[tbp]
\caption{Exchange energy (hartrees) per particle from three different
methods: XX-DFT $\left( \varepsilon _{x}^{XX}\right) $, LDA-DFT $\left(
\varepsilon _{x}^{LDA}\right) ,$ and VMC $\left( \varepsilon _{x}^{VMC},%
\text{ from Ref. 23}\right) $.}
\begin{tabular}{cccccc}
$q/k_{F}^{0}$ & $\varepsilon _{x}^{XX}$ & $\varepsilon _{x}^{LDA}$ & $%
\varepsilon _{x}^{VMC}$ & $\varepsilon _{x}^{XX}-\varepsilon _{x}^{VMC}$ & $%
\varepsilon _{x}^{XX}-\varepsilon _{x}^{LDA}$ \\ 
1.11 & -0.2911 & -0.2795 & -0.2930 & 0.0019 & -0.0116 \\ 
1.55 & -0.2739 & -0.2687 & -0.2756 & 0.0017 & -0.0052 \\ 
2.17 & -0.2469 & -0.2508 & -0.2534 & 0.0065 & +0.0039
\end{tabular}
\end{table}


\begin{references}
\bibitem{units}  Hartree (atomic) units are used in this work: $a_{0}=\hbar
^{2}/me^{2}\simeq 0.529$ $\AA$ is the unit of lenght, $\hbar
^{2}/ma_{0}^{2}\simeq 27.2$ eV is the unit of energy.
\bibitem{chemical}  In the present work all results are for a fixed number
of particles. From the numerical point of view, however, it is more
convenient to treat the model in Fig. 1 as an open system in contact with a
particle bath at chemical potential $\mu .$ To satisfy the constraint of
constant number of particles, the chemical potential $\mu $ is then adjusted
at each iteration of the self-consistent calculation loop.
\bibitem{3d}  Two subbands are occupied in the range $1\lesssim
q/k_{F}^{0}\lesssim 3;$ moving towards the long wavelength limit, the number
of occupied subbands progressively increases from two to eight in the range $%
0.5\lesssim q/k_{F}^{0}\lesssim 1.$
\end{references}
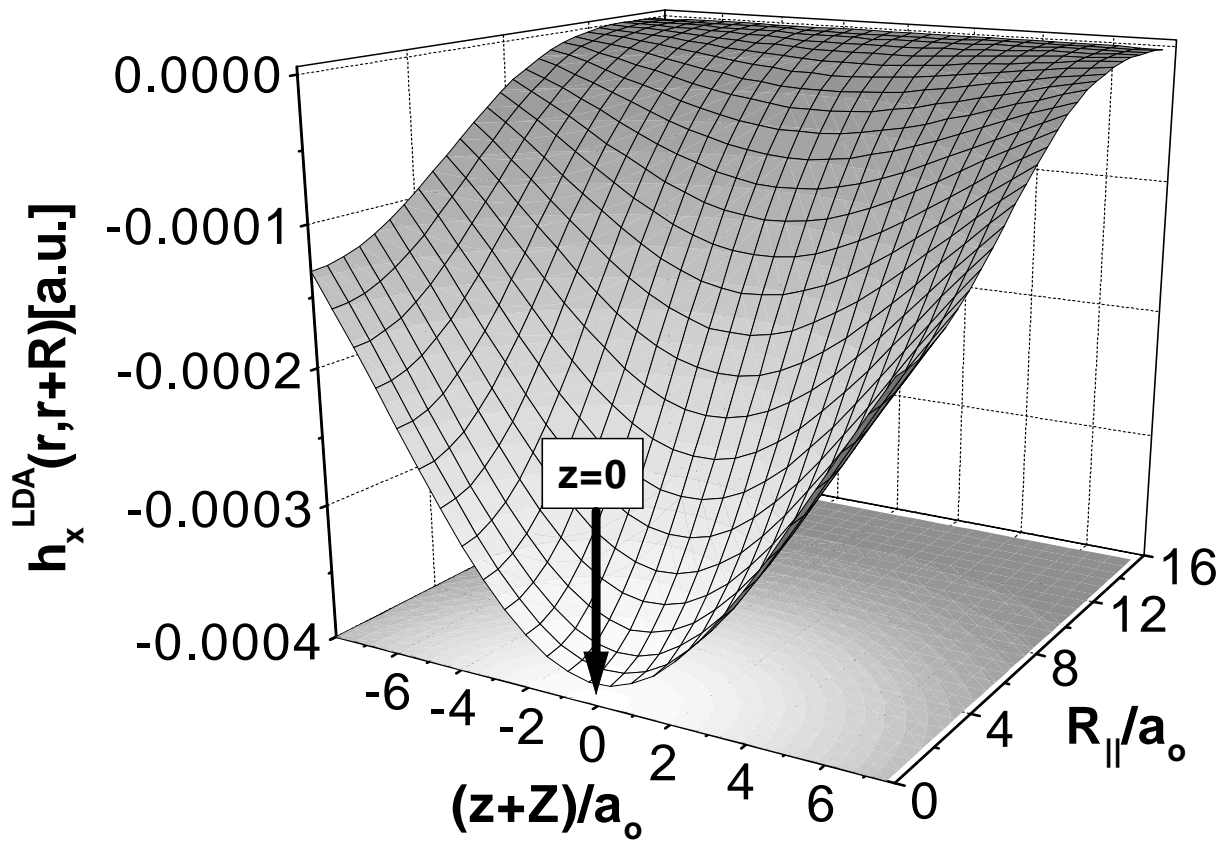
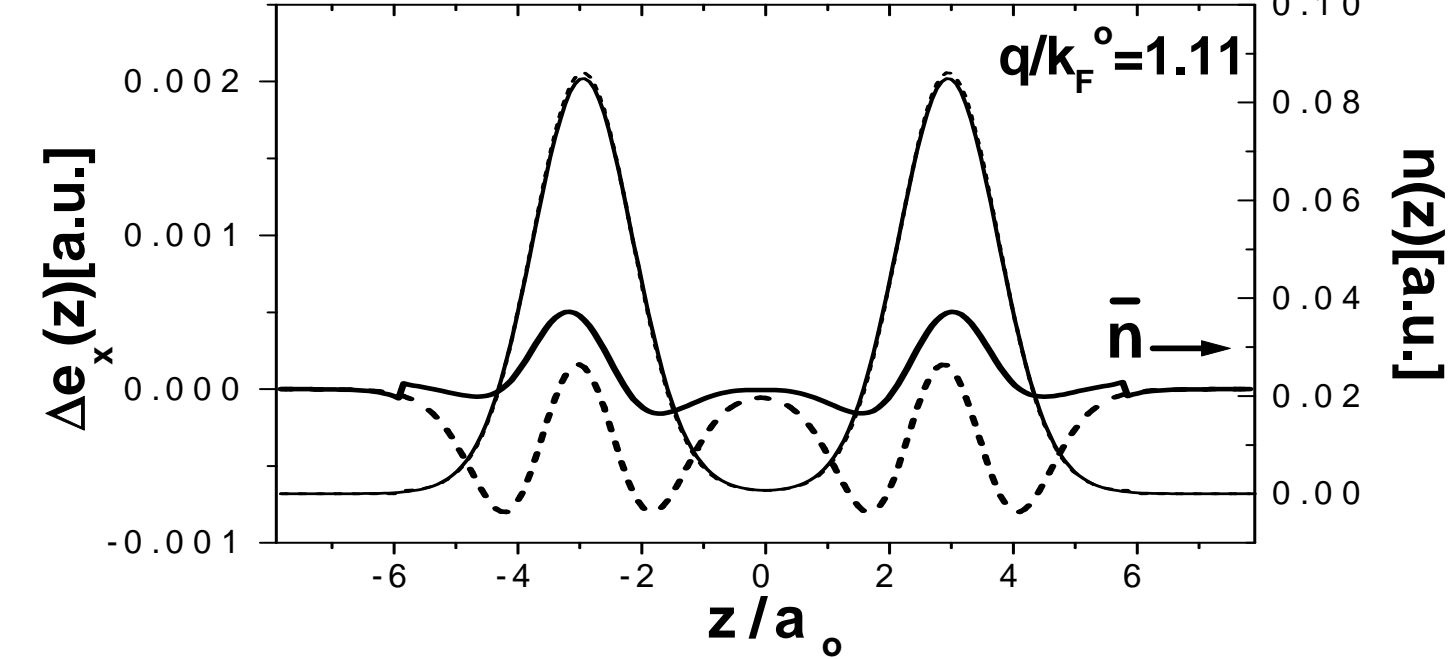
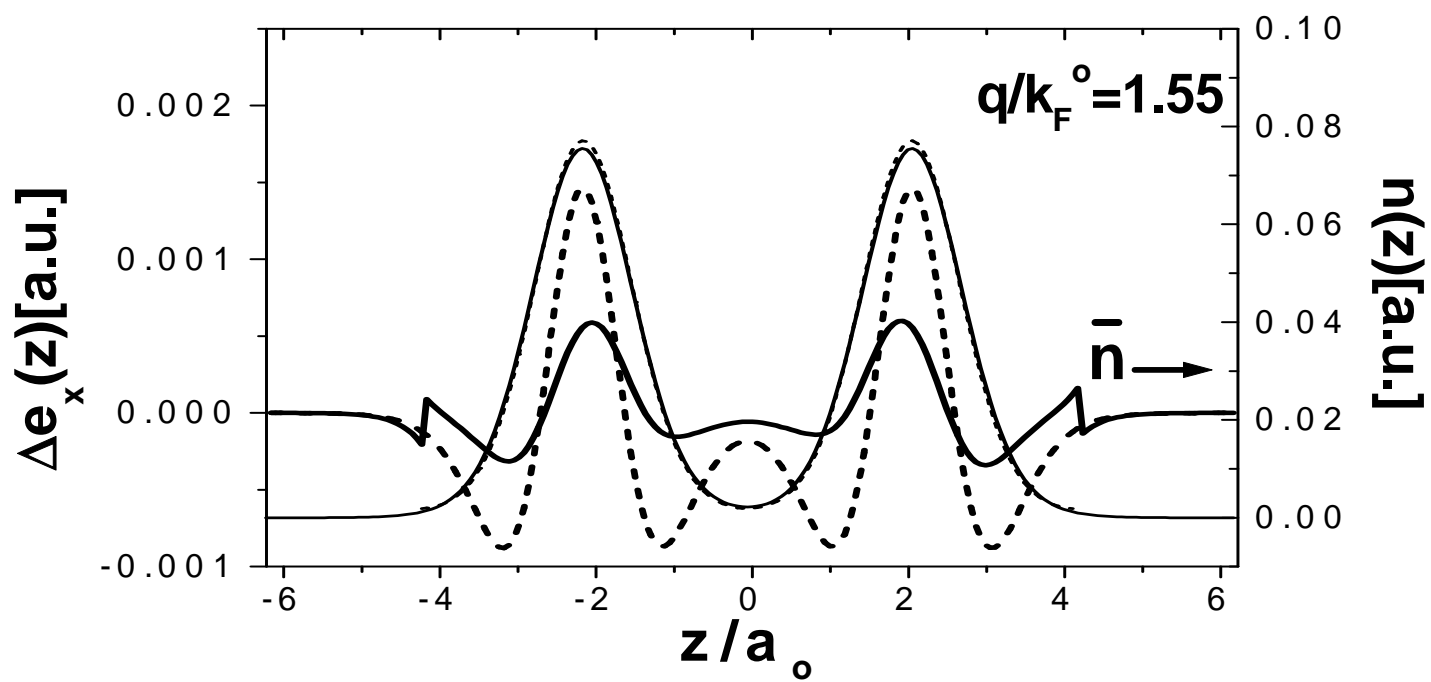
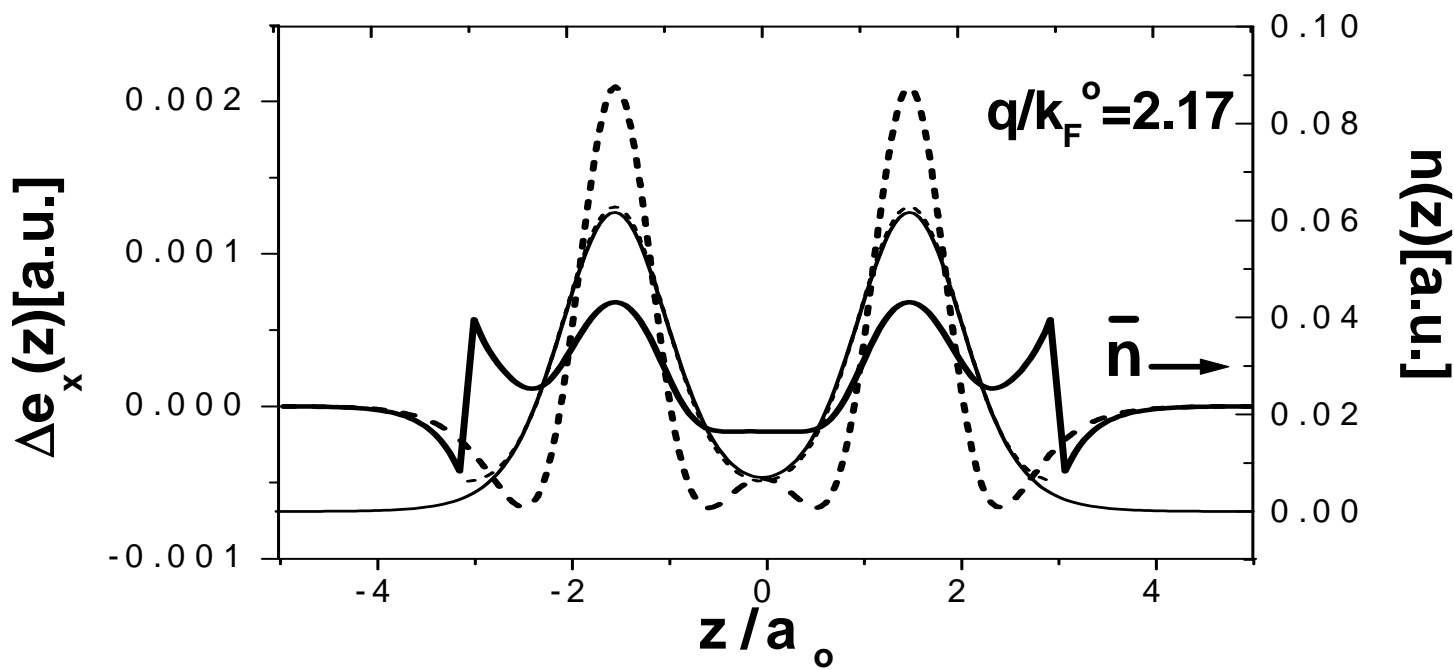
\end{document}